\documentclass[twocolumn,showpacs,preprintnumbers,amsmath,amssymb,prb]{revtex4-1}
\usepackage{graphicx}
\usepackage{dcolumn}
\usepackage{bm}
\usepackage{float}
\usepackage{color}

\begin{document}

\title{Inelastic X-ray scattering from valence electrons 
near absorption edges \\
of FeTe and TiSe$_2$}
\author{B. Barbiellini$^1$, 
J. N. Hancock$^{2,3}$, 
C. Monney$^4$, 
Y. Joly$^{5,6}$,
G. Ghiringhelli$^7$,
L. Braicovich$^7$,
T. Schmitt$^8$}
\affiliation{
$^1$Physics Department, Northeastern University, Boston, Massachusetts
02115, USA\\
$^2$Department of Physics, University of Connecticut, 
Storrs, Connecticut 06269, USA\\
$^3$Institute for Materials Science, 97 N. Eagleville Rd., 
Storrs, Connecticut 06269, USA\\ 
$^4$ Fritz-Haber-Insitut der Max-Planck-Gesellschaft, 
Faradayweg 4-6
D-14195 Berlin, Germany\\
$^5$ Universit\'e Grenoble Alpes, Institut NEEL, F-38000 Grenoble, France\\
$^6$ CNRS, Institut NEEL, F-38042 Grenoble, France\\
$^7$ CNR/SPIN and Dipartimento di Fisica, 
Politecnico di Milano, 
Piazza Leonardo da Vinci 32, 20133 Milano, Italy, \\
$^8$ Paul Scherrer Institut, CH-5232 Villigen PSI, Switzerland
}

\date{\today}

\begin{abstract}
We study resonant inelastic x-ray scattering (RIXS) peaks 
corresponding to low energy particle-hole excited states 
of metallic FeTe and  semi-metallic TiSe$_2$
for photon incident energy tuned near the
$L_{3}$ absorption edge of Fe and Ti respectively.  
We show that the cross section amplitudes are well 
described within a renormalization group theory where
the effect of the core electrons is captured by effective
dielectric functions expressed in terms of the
the atomic scattering parameters $f_1$ of Fe and Ti.
This method can be used to extract the dynamical 
structure factor from experimental RIXS spectra 
in metallic systems.

\end{abstract}

\maketitle

\section{introduction}

Inelastic x-ray scattering (IXS) probes charge and spin
electronic excitations in condensed matter 
but the corresponding cross section is
generally small \cite{ixs}, especially
the magnetic IXS cross section \cite{mag_ixs}. 
By tuning the incident energy near 
an x-ray absorption edge a cross section enhancement can be observed 
\cite{rixs1,keijo,rixs2,glatzel,rixs3,rixs4,jude,rixs5}.
Work by Kramers, Heisenberg, Weiskopf and Wigner
provides the foundation for quantitative approaches \cite{agren} 
for Resonant IXS (RIXS).
Nevertheless, some small disagreements between
theory and experiment
recently observed both in liquids \cite{sun} 
and in solids \cite{berkeley,JasonCuBO}
suggest that the Kramers-Heisenberg (KH)
theory misses some inelastic processes, 
especially at incident energy 
tuned sligthly lower than the absorption threshold.

The KH expression based  
on nonrelativistic second-order perturbation 
theory for the interaction between light and matter describes well
the radiation reemitted by atoms after illumination by primary 
radiation (e.g. resonant X-ray emission) 
but cannot fully describe collision 
processes between weakly bound electrons and light-quanta \cite{lacki}. 
The scattering of an x-ray photon from
nearly-free electrons involves both energy and momentum conservation
of an almost isolated system formed by two particles, namely, the electron
and the photon.
The Thirring theorem states that the cross section of this Compton-like effect 
with all radiative corrections reduces 
in the non relativistic limit to the Thomson cross section 
(i.e. non relativistic limit 
of the Klein-Nishima cross section \cite{booklet}). 
The only effect of the vacuum or of the medium is to renormalize 
the Thomson cross section \cite{heringen}.

In order to explore the scattering of nearly-free electrons
by a photon tuned near the binding energy of a core level,
we have introduced a scheme based 
on the renormalization group (RG) theory \cite{bn} able 
to cast the complicated RIXS problem to a much simpler IXS problem
in an effective medium described by an effective dielectric 
function $\epsilon$, 
which integrates out the effect of the core electrons.
In the RG approach, a differential equation describing the 
evolution of the coupling constant 
with screening yields the renormalization factor $\eta$ for the 
effective Thomson cross section in the polarizable medium.
This method effectively allows one to calculate the incident photon energy 
dependence of the RIXS cross section in the case of a metallic system.
Recently, Haverkort \cite{maurits} has shown that effective models 
based on the absorption spectra can give a good description of 
RIXS, although his approach was applied to spin excitations 
in strongly correlated systems rather than charge excitation 
in metallic systems as in the present case.

The present work studies high resolution
measurements \cite{jason,claude} near the Fe $L_{3}$ edge 
of Fe$_{1.087}$Te  
and the Ti $L_{3}$ edge of TiSe$_2$ by considering 
low energy particle-hole 
scattering peaks as a function of incident photon energy.
Previous RIXS studies at the $L_{3}$ edge of Cu and Ni
performed at lower resolution have neglected 
momentum conservation \cite{magnuson,winnick}.
Based on the changes in intensity of these peaks, we show 
that the resonant scattering cross section can be described via 
a renormalization factor $\eta$ 
multiplying the non-resonant IXS cross section.
In principle, the knowledge of the dynamical structure factor permits us to extract band structure information from measured dispersing electron-hole excitations. It also allows reconstructions of the density propagator of a system \cite{reed}, which yields the time dependent linear response of the system to a point perturbation. Therefore, momentum dependent RIXS can become a unique window for visualizing the dynamics of weakly bound electrons in condensed matter.

An outline of this paper is as follows. Section II summarizes our model.
In Sec. III, we present the methods for the electronic structure calculations and 
for the RIXS experiments. The results of the calculations are presented and 
compared with experimental results in Sec. IV, and the conclusions are given in Sec. V.

\section{Model}
In our model, the low-energy-loss scattering leads 
to particle-hole excited states of the valence electron system 
described by the double differential scattering cross-section,
which is the product of  the effective Thomson scattering cross
section $d\sigma_{Th}^*/d \Omega$ and 
of the dynamical structure factor $S(\textbf{q},\omega)$: 
\begin{equation}
\frac{d^2 \sigma}{d \Omega d \omega}= 
\frac{d\sigma_{Th}^*}{d \Omega} 
S(\textbf{q},\omega).
\label{eqkn}
\end{equation}
For particle-hole excitations 
the dynamical structure factor 
is given by \cite{ray}
\begin{equation}
S(\textbf{q},\omega)=
\int_{-\omega}^{0} \frac{dE}{2\pi} 
\int\frac{d^{3}p}{(2\pi)^{3}} 
A(\textbf{p},E)
A(\textbf{p}+\textbf{q},E+\omega),
\label{eqsqw}
\end{equation}
where $(\textbf{q},\omega)$ are the momentum and energy transfer
$A(\textbf{p},E)=(1/\pi) \mbox{Im}[G(\textbf{p},E)]$ is 
the imaginary part of the electron 
GreenÕs function $G(\textbf{p},E)$.
The spectral function $A(\textbf{p},E)$
can be also expressed in terms of the 
Dyson orbitals $g_{\nu}$ as follows:
\begin{equation}
A(\textbf{p},\omega)=\sum_{\nu} |g_{\nu}(\textbf{p})|^2
A_{\nu}(\omega),
\label{eqdyson1}
\end{equation} 
and
\begin{equation}
A_{\nu}(\omega)=
\frac{\gamma}{\pi[(\omega-\epsilon_{\nu})^{2}+\gamma^{2}]},
\label{eqA}
\end{equation}
where $\epsilon_{\nu}$ is the excitation energy
of the Dyson orbital $g_{\nu}(\textbf{r})$ 
and $\gamma$ is an inverse lifetime.
In the present case, the Dyson orbitals are given 
by Bloch wave functions, where $\nu=({\textbf{k},n})$ is given 
by the Bloch wave vector $\textbf{k}$ and by 
the energy band index $n$. Therefore one can write
\begin{equation}
g_{\textbf{k},n}(\textbf{r})=\exp(i\textbf{k}\cdot\textbf{r})
\sum_{\textbf{G}} C_{\textbf{G}}^{\textbf{k},n}\exp(-i\textbf{G}\cdot\textbf{r}),
\label{eqbloch1}
\end{equation}
whose momentum density is
\begin{equation}
|g_{\textbf{k},n}(\textbf{p})|^2=
\sum_{\textbf{G}}\delta(\textbf{p-k+G})~
|C_{\textbf{G}}^{\textbf{k},n}|^2.
\label{eqbloch2}
\end{equation}
The Fourier coefficients $C_{\textbf{G}}^{\textbf{k},n}$
of the periodic part of $g_{\textbf{k},n}(\textbf{r})$
are labeled by the reciprocal vectors $\textbf{G}$. 
Therefore, the spectral functions enforce the energy
conservation near the Fermi level and the momentum
conservation in the first Brillouin zone. 
 
Because of the proximity of the $L_{3}$ absorption 
threshold, the incoming photon generates a set of virtual
intermediate states involving a $2p$ core hole
and a corresponding electron excited in a $3d$ state which
can be described by an effective dielectric function $\epsilon$
experienced by the valence electrons.  
When the scattering process is completed, it leaves behind
low-lying valence excited states conserving energy and crystal
momentum and the Thomson cross section $\sigma_{Th}$ 
of this resonant scattering is increased by the factor
$\eta$ compared to the non resonant
cross section.
The solution of the renormalization group equation 
gives \cite{bn} 
\begin{equation}
\eta=\exp[\frac{2}{3 \alpha}(\epsilon/\epsilon_0 -1)],
\end{equation}
where 
$\alpha$ is the fine structure constant,
$\epsilon$ is the real part of an effective
dielectric function for the valence electrons 
as a function of 
energy $\omega$ and $\epsilon_0$
is the dielectric constant in vacuum.
One can connect $\epsilon$ to the 
atomic scattering parameter $f_1$ 
through the formula
\begin{equation}
\frac{\epsilon}{\epsilon_0}=1- (\frac{\omega_P}{\omega})^2 f_1,
\end{equation}
where $\omega_P$ is the plasma frequency
for the valence electron gas at 
the Fermi level, which can be written as
\begin{equation}
\omega_P=47.1 ~r_s^{-3/2}~\mbox{eV},
\end{equation}
where $r_s$ is the 
radius containing a valence electron.
The parameter $f_1$ is related to the 
absorption coefficient $\mu(E)$ via
the Kramers-Kr\"onig transform \cite{prange},
\begin{equation}
f_1(E)= Z+\frac{2}{\pi}
        \int_{0}^{\infty} 
        \frac{\omega_P^2\mu(\omega)}
             {E^2-\omega^2} d\omega,
\end{equation}
where $Z$ is the atomic charge.
The cross section enhancement occurs 
only if $f_1$ is negative.
The parameter $f_1$ can be also 
determined by reflectivity experiments
\cite{sacchi,stone}.
The enhancement term is therefore given by
\begin{equation}\label{eqn_eta}
\eta=\exp(-\frac{2}{3 \alpha}(\frac{\omega_P}{\omega})^2 f_1), 
\end{equation}
and it can be expanded in a power series of 
the parameter $f_1$ 
which contains a Lorentz oscillator \cite{booklet} for 
the core 2$p$ electrons appearing also in the KH treatement
of RIXS.

\section{Computational and experimental methods}
We have calculated $f_1$ from first-principles
using the program FDMNES \cite{joly}
within the Time Dependent Density Functional Theory (TDDFT)
\cite{TDDFT}.
First, we have performed local density approximation
self-consistent calculations 
for FeTe and TiSe$_2$,
and then we computed
$f_1$ at the Fe $L_3$ edge and 
at the Ti $L_3$ edge using TDDFT.
The calculations for Ti are performed 
for the $L_3$ and the $L_2$
edges since the $L_2$ edge 
influences the $f_1$ value
near the $L_3$ edge.
The influence of the $L_2$ edge
can be neglected in Fe since 
the separation between the $L_2$
and the $L_3$ edge is larger in this case.
The crystal structure of FeTe is tetragonal with point group
$P4/nmm$ and lattice constants
$a=b=3.8215$ \AA, $c=6.2695$ \AA\  while the crystal structure of
TiSe$_2$ has a point group $P-3m1$ with lattice constants
$a=b=3.54$ \AA\  
and $c=6.008$ \AA.

The experimental RIXS spectra considered in 
this paper were taken at the ADRESS beam line \cite{beamline}, 
of the Swiss Light Source, Paul Scherrer Institut, 
using the SAXES spectrometer \cite{SAXES}. 
A scattering angle of $130^\circ$ was used, and the samples were measured 
at an incidence angle of $65^\circ$, using $\sigma-$polarized light. 
At the Ti $L_3$ edge, 
the combined energy resolution was 90 meV. 
TiSe$_2$ samples were measured at 16K. At the 
Fe  $L_3$ edge, the combined energy resolution was 73 meV. 
The FeTe samples were measured near 20K.
Our results showing the inelastic peak height dependence 
on the incident photon energy were 
obtained by integrating the RIXS intensity between 0.1 
and 4.0 eV energy loss for FeTe and between 0.1 and 0.6 eV energy loss for TiSe$_2$, after having subtracted 
a linear background and the elastic line.
Our RIXS spectra were also corrected for 
self-absorption effects, following the method used in 
Ref.~\onlinecite{JasonCuBO}.

\section{Results}

\begin{figure}
\includegraphics[width=8.0cm]{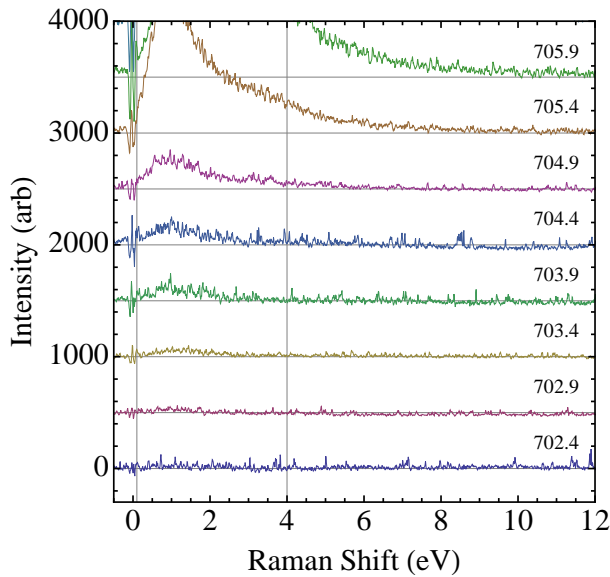}
\caption{(Color online) Focus on the RIXS subthreshold 
feature of Fe$_{1.087}$Te.
We concentrate on the 0.9 eV energy loss; however, we consider
the area below 0.1 and 4.0 eV to estimate the enhancement
of the peak.
This spectrum has been corrected for 
self-absorption effects, following the method used in 
Ref.~\onlinecite{JasonCuBO}. 
The elastic line has also been removed.} 
\label{fete}
\end{figure}

We show in Fig.~\ref{fete} a resonant 
feature below the Fe $L_{3}$ threshold 
of Fe$_{1.087}$Te.
The subthreshold peak energy locks 
to a constant value of 0.9 eV while the RIXS 
line shape preserves its shape.
Interestingly, the momentum transfer of this peak is $q=0.51$ \AA~, 
and the energy transfer of about $0.9$ eV
corresponds to the recoil energy of an electron
or a hole $q^2/(2m^*)$ with an effective mass $m^*\sim 1$.
This feature is also consistent with excitations associated with 
bands crossing the Fermi level around the $\Gamma$ point 
forming small hole pockets 
and around the $M$ point
giving small electron pockets \cite{alaska}.
Similar Fermi surfaces are also found 
in the LaO$_{1-x}$F$_x$FeAs 
iron pnictide compounds \cite{ray1,ray2}.

\begin{figure}
\includegraphics[width=8.0cm]{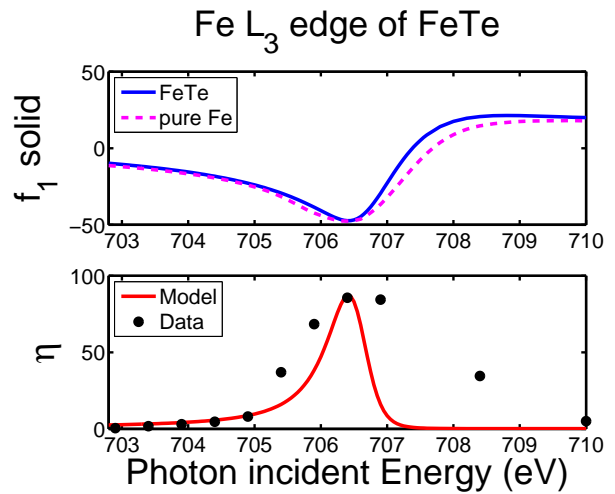}
\caption{(Color online) Fe $L_{3}$ edge of Fe$_{1.087}$Te:
In the top frame the solid curve represents the model
for $f_1$ as a function of the incident energy.
The dashed curve represents $f_1$ data 
from total electron yield experiments in pure iron.\cite{sebastian}
The label "$f_1$ solid" emphasizes that 
the parameter $f_1$ for the solid is different from
the one calculated for free atoms. 
In the bottom frame the black circles and the 
solid curve are inelastic peak height
and the model with $f_1$ from FDMNES 
respectively, plotted against incident energy.}
\label{f1}
\end{figure}

The calculated $f_1$ 
as a function of incident energy is shown 
in the top frame of Fig.~\ref{f1} together
with  $f_1$ data 
from total electron yield experiments 
in pure iron \cite{sebastian}.
The agreement between the two curves is quite
impressive. Therefore, the behavior of $f_1$
is very similar in Fe$_{1.087}$Te
and in pure iron.
This result confirms that the 
x-ray absorption for materials 
related to iron pnictides is
qualitatively similar to Fe metal \cite{pnictides}
but also very different from Fe ions in the
Li$_x$FePO$_4$ compound for the Li-battery 
cathodes \cite{rixs_lfp}, where multiplet calculations must be included 
\cite{jacs}.
In the bottom frame of Fig.~\ref{f1} 
the inelastic peak height 
is plotted against
incident energy 
together with the theoretical enhancement
$\eta$. We have used the value
$r_s= 1.60$ a.u., which 
is consistent with
the plasmon energy losses
associated with the Fe 1$s$ core 
level \cite{watts}. 
The corresponding 
plasma energy 
of the Fe 3$d$ shell is 23.3 eV.
The overall agreement between the experimental data 
and $\eta$ is good below the threshold energy of 705 eV, 
and it remains reasonable after this threshold despite 
the fact that the present theory does not include the 
resonant x-ray emission described by KH.
The enhancement factor $\eta$ is very large,
and it approaches a value as large as 87.

\begin{figure}
\includegraphics[width=8.0cm]{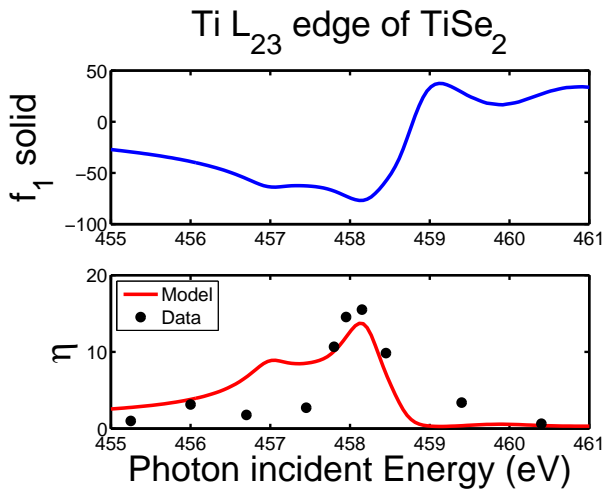}
\caption{(Color online) Same as Fig.~\ref{f1} 
for the Ti $L_{23}$ edge of TiSe$_2$ (for an integrated 
energy loss region between 0.1 eV and 0.6 eV).} 
\label{f2}
\end{figure}

In the case of TiSe$_2$, 
the choice of the parameter $r_s=3.0$
yields a screening length $\lambda=\sqrt{1.56 r_s}= 2.16$ a.u. 
within the Thomas Fermi model \cite{bba89}
which is consistent with the measured plasma frequency 
\cite{hasan,claudenjp}. 
This larger $r_s$ value gives a much smaller 
enhancement $\eta$ 
compared to the FeTe case, which reaches
a factor of 14 at most.
The FDMNES calculation of $f_1$
yields amplitudes of about $-77$ around 
an energy of $458$ eV.
In the present case,
well defined particle-hole features given by $S({\bf q},\omega)$ 
dominate the fluorescent contributions
in agreement with the findings by Monney {\em et al.} \cite{claude}.
As mentioned above, the $L_2$ and $L_3$ edges of Ti almost overlap. 
Therefore, one must perform a calculation for both edges 
at the same time in order to have 
an accurate description of the amplitude $f_1$.
The TDDFT must also be applied in order to describe many-electron 
effects which modify the ratio of the $L_2$ and $L_3$
contributions. However, as shown in the 
bottom of Fig.~\ref{f2} the agreement with the experiment 
is still not perfect, despite the modelÕs ability 
to capture the main amplitude in the RIXS signal.

\begin{figure}
\includegraphics[width=8.0cm]{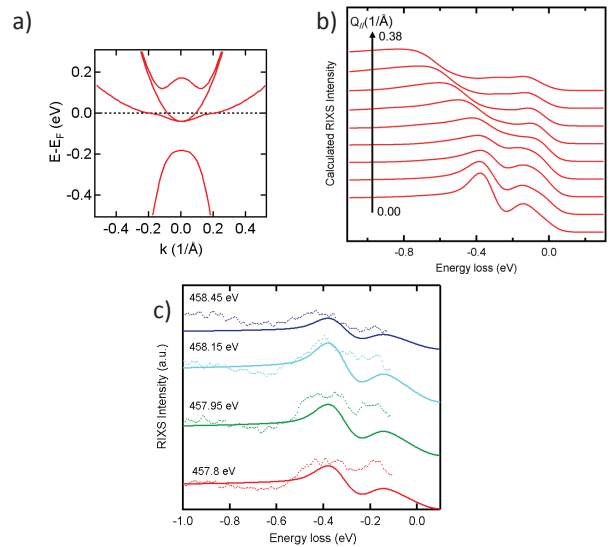}
\caption{(Color online) Model for the RIXS in TiSe$_2$:
(a) Band structure given in Ref.\onlinecite{claude},
(b) Dispersion of $S(\textbf{q},\omega)$
(c) Calculated RIXS spectrum at $q_{||}=0$ (continuous lines), 
together with experimental RIXS spectra 
(dotted lines) of TiSe$_2$ for selected incident photon 
energies, written on the left (at 16K and for $q_{||}\sim 0$) 
from Ref.\onlinecite{claude} (spectra plotted up to the elastic line).} 
\label{frixs}
\end{figure}

To illustrate our approach, we calculate 
the dynamical structure factor $S(\textbf{q},\omega)$
for four bands near the Fermi level, 
as performed previously \cite{claude} 
to understand electron-hole excitation 
dispersions in the charge density wave phase of TiSe$_2$.
These four bands are depicted in Fig.~\ref{frixs} (a)
and correspond to a model band structure of the charge density wave in TiSe$_2$, inspired by angle-resolved photoemission spectroscopy experiments. Transitions between the lowest occupied band and the highest unoccupied band give rise to the dispersive peak at high energy losses in RIXS, while transitions between the partially occupied bands and the highest unoccupied one give rise to the dispersive peak at low energy losses.
The corresponding spectral functions have been broadened 
using a Gaussian function to account for the experimental resolution. 
Figure~\ref{frixs} (b) shows the resulting RIXS spectra as a function 
of the transferred momentum of light 
projected on the surface plane of the sample, $q_{||}$.
These spectra are given by an energy-loss- and momentum-resolved 
convolution of the occupied and unoccupied band structure, which
approximates Eq.~\ref{eqsqw}.
Reference~\onlinecite{claude} shows that 
such a convolution compares very well with the measured $q$-dependent RIXS spectra 
for an incident photon energy maximizing the cross section. 
According to Eq.~\ref{eqkn}, the cross section 
can be obtained using the enhancement term $\eta$ given
by Eq.~\ref{eqn_eta}.
Figure~\ref{frixs} (c) shows selected 
calculated RIXS spectra (solid lines) 
obtained with the dynamical structure factor 
function at $q_{||}=0$ and multiplied by the enhancement 
factor $\eta$ for TiSe$_2$.
For comparison, we show in the same panel 
the corresponding experimental 
RIXS spectra (dotted lines) of TiSe$_2$ 
at different incident photon energies, 
(see Ref.~\onlinecite{claude} for more details).
Thus, this example illustrates 
the present RIXS model 
for particle-hole excitations in metallic systems.
This scheme factorizes the RIXS signal  
in an incident-energy dependent part  
describing the resonance behavior multiplied 
by an energy-loss and momentum 
dependent term for 
the excitation dispersion.

\section{Conclusions}
We propose, by using a renormalization 
group approach,
to simplify the description of RIXS for 
the particle-hole excitations 
in metallic systems.
For this purpose, we connect $S({\bf q},\omega)$ 
to an effective scattering cross section 
where the amplification factor $\eta$ is directly related to 
the atomic scattering parameter $f_1(\omega$), 
which is strongly modulated across resonant absorption edges.
For metallic Fe and FeTe, the TDDFT calculation of $f_1(\omega$),  
works very well while for  
TiSe$_2$ one can expect some discrepancy
between theory and experiment
since these calculations usually 
can reproduce rather well the absorption cross section 
(related to $f_1$) for metals, 
but not the absorption in ionic or correlated systems. 
Nevertheless, we find that the model
is still able  
to capture the overall amplitude of the RIXS signal
in TiSe$_2$. 
The reason for the discrepancies between model 
and experiment above the absorption thresholds is 
explained by the fact that the present model does 
not include the momentum independent resonant emission described by KH.
An important question is what can we learn from the present approach?
The band structure calculations for TiSe$_2$ 
shown by Monney {\em et al.}~\cite{claude} already 
reproduced many interesting features of the RIXS spectral shape, 
including the dispersion embedded in 
$S({\bf q},\omega)$. However, the missing part of this previous work 
was the amplitude of the RIXS process involving
both energy and momentum conservation,
which can now be brought in by our present model.
Thus, the most interesting aspect of the present model 
is a method to extract $S({\bf q},\omega)$ in metallic systems 
in future ${\bf q}$-dependent RIXS experiments.

\begin{acknowledgments}
We thank Sebastian Macke for sending us $f_1$ data for iron.
B.B. is supported by the US Department of Energy (USDOE) 
Contract No. DE-FG0207ER46352.
C.M. and T.S. acknowledge support from the 
Swiss National Science Foundation (SNSF) and 
its National Centre of Competence in Research MaNEP.
C.M. acknowledges support by the Swiss National Science Foundation 
(under Grant No. PA00P2\textunderscore142054) 
and from the Alexander von Humboldt Foundation.
We benefited from computer time from 
Northeastern UniversityÕs Advanced Scientific Computation Center (ASCC) 
and USDOEÕs NERSC supercomputing center.
\end{acknowledgments}

\end{document}